\documentstyle[12pt,epsfig]{article}
\newlength{\dinwidth}
\newlength{\dinmargin}
\newcommand{\ba}{\begin{array}}
\newcommand{\ea}{\end{array}}
\newcommand{\beq}{\begin{equation}}
\newcommand{\eeq}{\end{equation}}
\newcommand{\bea}{\begin{eqnarray}}
\newcommand{\eea}{\end{eqnarray}}

\def\S{{\bf S}}

\def\bce{\begin{center}}
\def\ece{\end{center}}

\def\nonu{\nonumber}

\def\pa{\partial}
\def\al{\alpha}
\def\be{\beta}
\def\ga{\gamma}

\def\De{\Delta}

\def\si{\sigma} \def\Si{\Sigma}

\def\S{{\bf S}}

\begin{document}
\thispagestyle{empty}
\addtocounter{page}{-1}
\begin{flushright}
{\tt hep-th/0206176}\\
\end{flushright}
\vspace*{1.3cm}
\centerline{\Large \bf  Penrose Limit of $AdS_4 \times N^{0,1,0}$
and  ${\cal N}=3$ Gauge Theory }
\vspace*{1.5cm} 
\centerline{ \bf Changhyun Ahn
}
\vspace*{1.0cm}
\centerline{\it 
Department of Physics,}
\vskip0.3cm 
\centerline{  \it Kyungpook National University,}
\vskip0.3cm
\centerline{  \it  Taegu 702-701, Korea }
\vspace*{0.3cm}
\centerline{\tt ahn@knu.ac.kr 
}
\vskip2cm
\centerline{\bf  abstract}
\vspace*{0.5cm}

We consider M-theory on $AdS_4 \times N^{0,1,0}$ where
$N^{0,1,0}= (SU(3) \times SU(2))/(SU(2) \times U(1))$. 
We review a Penrose limit of $AdS_4 \times N^{0,1,0}$ that 
provides the pp-wave geometry of $AdS_4 \times \S^7$. 
There exists a subsector of 
three dimensional ${\cal N}=3$ dual gauge theory, by taking
both the conformal dimension and $R$-charge large with the finiteness 
of their difference, which 
has enhanced ${\cal N}=8$ maximal supersymmetry.
We identify operators in the ${\cal N}=3$ gauge theory with
supergravity KK excitations in the pp-wave geometry and 
describe how the ${\cal N}=2$ 
gauge theory operators originating from both ${\cal N}=3$ short vector
multiplet and ${\cal N}=3$ long gravitino multiplet fall into 
${\cal N}=8$ supermultiplets.

\vspace*{4.0cm}

\baselineskip=18pt\newpage
\section{Introduction}

The large $N$ limit of a subsector of $d=4, {\cal N}=4$ $SU(N)$
supersymmetric gauge theory is dual \cite{bmn} to type IIB string theory in 
the
pp-wave background \cite{bfhpetal,bfhpetal1}. 
This subspace of the gauge theory
is described by string theory in the pp-wave background. By taking
a scale limit of the geometry near a null geodesic in $AdS_5 \times \S^5$,
it gives rise to the appropriate subspace of the gauge theory. The operators
with large $R$-charge in the subsector of ${\cal N}=4$ $SU(N)$ gauge theory
were identified with the stringy states in the pp-wave background.
It was found that the Penrose limit of $AdS_5 \times T^{1,1}$ provides
pp-wave geometry of $AdS_5 \times \S^5$ \cite{ikm,go,zs}. 
Using AdS/CFT correspondence, one
can identify gauge theory operators with large $R$-charge with the stringy
excitations in the pp-wave geometry. Moreover, the maximal ${\cal N}=4$
multiplet structure hidden in the ${\cal N}=1$ gauge theory can be 
predicted from both a chiral operator and semi-conserved operator with large 
$R$-charge.  
   
It is natural to think about the subsector of ${\cal N}=2$
gauge theory in $d=3$ in the context of $AdS_4 \times {\bf X}^7$ where
${\bf X}^7$ is an Einstein seven manifold. 
Recently the operators with large $R$-charge in the boundary field 
theory were obtained from the complete spectrums of 11-dimensional KK 
compactifications on $AdS_4 \times Q^{1,1,1}$ \cite{ahn02},
$AdS_4 \times M^{1,1,1}$ \cite{ahn02-1} and $AdS_4 \times V_{5,2}$ 
\cite{ahn02-2}
in pp-wave limit. In old days,
all the supersymmetric ${\cal N}=2$ 
homogeneous manifolds were classified in \cite{crw}.
There exist only three ${\cal N}=2$ theories and they are $Q^{1,1,1}, 
M^{1,1,1}$ and $V_{5,2}$. The isometry of these manifolds corresponds to
the global symmetry of the dual SCFT including $U(1)_R$ symmetry of
${\cal N}=2$ supersymmetry.      

In this paper, we consider a similar duality
that is present between a certain three dimensional ${\cal N}=3$
gauge theory and 11-dimensional supergravity theory
in a pp-wave background with the same spirit as in 
\cite{ahn02,ahn02-1,ahn02-2,go,ikm,zs}.
This is a continuation of previous considerations 
\cite{ahn02,ahn02-1,ahn02-2}. 
We describe this duality by taking a scaling limit of the duality
between 11-dimensional supergravity on $AdS_4 \times N^{0,1,0}$
where $N^{0,1,0}$ was found in \cite{cr} and three dimensional
${\cal N}=3$ superconformal field theory. The boundary theory in terms of
${\cal N}=2$ superfields is a gauge theory
with gauge group $SU(N) \times SU(N)$ with chiral fields
$U^i$ transforming in the $({\bf N}, {\bf \overline{N}})$ 
color representation and ${\bf 3}$ under the flavor group
$SU(3)$ and $V_i$  transforming in the $({\bf \overline{N}}, {\bf N})$ 
color representation and  ${\bf \overline{3}}$ under the flavor group. 
The complete analysis on the spectrum of
$AdS_4 \times N^{0,1,0}$ was found in \cite{termonia,freetal}(See also
\cite{cast}). 
This gives the theory that lives on $N$ M2-branes 
at the conical singularity of a Calabi-Yau four-fold. 
The scaling limit is obtained by considering the geometry near a null 
geodesic carrying large angular momentum in the $SO(3)_R=SU(2)_R$
isometry of the $N^{0,1,0}$ space which is dual to the $SU(2)_R$ R-symmetry
in the ${\cal N}=3$ superconformal field theory.

In section 2, we review the scaling limit around
a null geodesic in $AdS_4 \times N^{0,1,0}$ from the explicit
metric of $N^{0,1,0}$ 
and obtain a pp-wave 
background \cite{gns}.
In section 3, we identify supergravity excitations in the Penrose limit
with gauge theory operators. What we observed is the presence of
a semi-conserved field in the ${\cal N}=3$ 
long gravitino multiplet  propagating
in the $AdS_4$ bulk. 
In section 4, we summarize our results.

\section{Penrose Limit of $AdS_4 \times N^{0,1,0}$}

Let us start with the supergravity solution dual to
the ${\cal N}=3$ superconformal field theory \cite{billoetal}.
By putting a large number of $N$ coincident M2-branes at the
conifold singularity and taking the near horizon limit,
the metric becomes  that \cite{pagepope} of 
$AdS_4 \times N^{0,1,0}$
\bea
ds_{11}^2 =ds_{AdS_4}^2 +ds_{N^{0,1,0}}^2,
\label{metric}
\eea
where
\bea
ds_{AdS_4}^2 & = & L^2 \left( -\cosh^2 \rho \; d t^2 +d \rho^2 + 
\sinh^2 \rho \; d \Omega_2^2 \right),
\nonu \\
ds_{N^{0,1,0}}^2 & = & 
\frac{L^2}{2}  \left[ \left( \Si_1 -\cos \zeta \si_1 
\right)^2 + \left( \Si_2 -\cos \zeta \si_2\right)^2+
\left( \Si_3 -\frac{1}{2} \left( 1+ \cos^2 \zeta \right) 
\si_3 \right)^2 \right]  \nonu \\
&& +  L^2 \left[ d \zeta^2 + \frac{1}{4} \sin^2 \zeta \left( \si_1^2 +
\si_2^2 +\cos^2 \zeta  \si_3^2 \right) \right]
\label{metric1}
\eea
where $ d \Omega_2$ is the volume form of a unit $\S^2$
and the curvature radius 
$L$ of $AdS_4$ is given by
$(2L)^6= 32 \pi^2 \ell_p^6 N$. 
Topologically $N^{0,1,0}$ 
is a nontrivial $SO(3)$ bundle
over ${\bf CP}^2 $. 
The coordinate $\zeta$  and $SU(2)$ one-forms $\si_i$ 
are the same as in the ${\bf CP}^2$ metric and the one-forms
$\Si_i$ are left-invariant forms on the manifold $SO(3)$.
The $SU(3) \times SU(2) $ isometry
 group of $N^{0,1,0}$ consists of $SU(3)
 $ global symmetry and $SO(3)_R$ symmetry of the dual 
superconformal field theory of \cite{billoetal}.
\footnote{
The three real one-forms $\si_i$ are given by 
with the ranges, $0 \leq \theta \leq  \pi,
0 \leq \phi,  \leq 2 \pi, 0 \leq \psi \leq 4 \pi$
\bea
\si_1 = \cos \psi d \theta + \sin \psi \sin \theta d \phi, \qquad
\si_2 = - \sin \psi d \theta + \cos \psi \sin \theta d \phi, \qquad
\si_3 =  d \psi + \cos \theta d \phi. \nonu
\nonu 
\eea
These one-forms are left-invariant under the $SU(2)$ group
and satisfy $SU(2)$ algebra.
Similarly the explicit form of left-invariant forms on the $SO(3)$
group manifold
in terms of angular variables are
\bea
\Si_1 = \cos \ga d \al + \sin \ga \sin \al d \be, \qquad
\Si_2 = - \sin \ga d \al + \cos \ga \sin \al d \be, \qquad
\Si_3 =  d \ga + \cos \al d \be.
\nonu
\eea
They also satisfy similar $SO(3)$ algebra.
Sometimes one uses the notation of $N(k,l)$ instead of $N^{pqr}$ due to the
redundancy of last integer $r$. The parameters are related by $k/l=
(3p+q)/(3p-q)$. Strictly speaking, 
the space $N^{0,1,0}$ we are dealing with in this paper
is $N^{0,1,0}_{I}$ which has ${\cal N}=(3,0)$ supersymmetry. 
There exists other type of the $N^{0,1,0}$ found by 
Page and Pope \cite{pagepope} and
denoted by $N^{0,1,0}_{II}$ that has
${\cal N}=(1,0)$ supersymmetry. In this case, the metric is also different 
from (\ref{metric1}). The squashing parameter in this case is $1/10$
not $1/2$ like as (\ref{metric1}).}

Let us make a scaling limit around a null geodesic in 
$AdS_4 \times N^{0,1,0}$. 
Let us introduce
coordinates which label the geodesic   
\bea
x^{+}  =  \frac{1}{2} \left[ t +\frac{1}{\sqrt{2}} \left( \ga + 
\be -\frac{1}{2} \psi - \frac{1}{2} \phi \right) \right], \qquad
x^{-}  =   \frac{L^2}{2} \left[ t -\frac{1}{\sqrt{2}} \left( \ga + 
 \be -\frac{1}{2} \psi - \frac{1}{2} \phi \right) \right],
\nonu
\eea
and make a scaling limit around $\rho=0=\theta  =\al$ and $\zeta = \pi/2$
in the above geometry (\ref{metric}).
By taking the limit $L \rightarrow \infty$ while 
rescaling the coordinates \cite{gns}
\bea
\rho=\frac{r}{L}, \qquad \zeta =\frac{\pi}{2}+\frac{  \zeta_1}{2 L}, \qquad
  \theta =\frac{ \zeta_2}{ L}, \qquad \al =
\frac{ \zeta_3}{ \sqrt{2} L}, \nonu
\eea
the Penrose limit of the  $AdS_4 \times N^{0,1,0}$ becomes \cite{gns}
\bea
ds_{11}^2 & = & -4 dx^{+} dx^{-} +
 \sum_{i=1}^3 \left( dr^i dr^i -r^i r^i dx^{+}
dx^{+} \right) + \frac{1}{4}
\sum_{i=1}^3 \left( d \zeta_i^2 +\zeta_i^2 d\phi_i^2
- \sqrt{2} \zeta_i^2 d\phi_i dx^{+} \right) \nonu \\
&= &  -4 dx^{+} dx^{-} +\sum_{i=1}^3 \left[ 
\left( dr^i dr^i -r^i r^i dx^{+}
dx^{+}  \right)
+ \frac{1}{4}
\left( d z_i d \bar{z}_i 
+  i  \sqrt{2} \left( \bar{z}_i d z_i - 
z_i d \bar{z}_i \right) dx^{+} \right) \right]
\label{ppwave}
\eea
where we define $\phi_1 = \frac{1}{2} \left(\psi +\phi \right),
\phi_2=-\phi, \phi_3 =\be$
and in the last line we introduce the 
 complex coordinates $z_i=\zeta_i e^{i \phi_i}$. 
Since the metric has a covariantly constant null Killing
vector $\pa / \pa_{x^{-}}$, it is also pp-wave metric.
Note that the pp-wave geometry (\ref{ppwave}) in the scaling limit reduces 
to
the maximally supersymmetric pp-wave solution of
$AdS_4 \times \S^7$ \cite{kow,op}
\bea
ds_{11}^2 = -4 d x^{+} d x^{-} +\sum_{i=1}^9  dr^i dr^i -
\left( \sum_{i=1}^3 r^i r^i  + \frac{1}{4}  \sum_{i=4}^9 
r^i r^i \right) dx^{+} dx^{+}. 
\nonu
\eea 
The supersymmetry enhancement in the Penrose limit
implies that a hidden ${\cal N}=8$ supersymmetry is present in the 
corresponding subsector of the dual ${\cal N}=3$ superconformal field 
theory.  In the next section, we provide precise description
of how to understand the excited states in the supergravity theory that 
corresponds to operators in the dual superconformal field theory. 

\section{Gauge Theory Spectrum}

The 11-dimensional supergravity theory in $ AdS_4 \times N^{0,1,0}$
is dual to the ${\cal N}=3$ gauge theory 
with gauge group
$SU(N) \times SU(N) $
coupled to a suitable set of hypermultiplets with Chern Simons interaction.
In ${\cal N}=2$ language, the field contents are given by   
two kinds of chiral fields $U^i, i=1, 2, 3$ transforming in the 
$\left({\bf N}, {\bf \overline{N}} \right) $  color representation and 
$V_j, j=1, 2, 3$ transforming in the 
$\left({\bf \overline{N}}, {\bf N} \right) $  
color representation  
\cite{billoetal} and
they transform as ${\bf 3}$ 
and ${\bf \overline{3}}$ under the $SU(3)$ global symmetry, respectively.
We identify states in the supergravity containing both short
and long multiplets with operators in the gauge theory.  
In each multiplet, we specify a $SU(3) $ representation 
\footnote{A representation of $SU(3)$ can be identified by a Young diagram
and when we denote the Dynkin label $(M_1, M_2)$ so that totally
we have $M_1+2M_2$ boxes, the dimensionality of
an irreducible representation is $N(M_1,M_2)=(1+M_1)(1+M_2)
(\frac{2+M_1+M_2}{2})$.  
Also an irreducible representation of $SU(2)$
can be described by a Young diagram with $2J$ boxes. Its dimensionality 
is $2J+1$.  }, conformal
weight and $SU(2)$ isospin in ${\cal N}=3$ superspace.

$\bullet$ {\bf ${\cal N}=3$ Massless(or ultrashort) multiplets} 
\cite{termonia,freetal,billoetal}

$1)$ One ${\cal N}=3$  massless graviton multiplet $\Theta^{\al}(x, 
\theta^{\pm}, \theta^0)$: 
$
{\bf 1}, \qquad
 \Delta=3/2, \qquad J=0
$ 

This massless graviton multiplet coresponds to the lowest($J=0$)
isospin ${\cal N}=3$ superfield. It is convenient to describe it in terms of
${\cal N}=2$  superfield because the structure of $OSp(2|4)$ multiplets
and its description as constrained superfield are known. 
An ${\cal N}=2$ stress-energy tensor superfield 
$T_{(\al \be)}(x, \theta^{\pm})$ corresponding to 
${\cal N}=2$ massless graviton multiplet 
satisfies
the equation  $D_{\al}^{\pm} T^{(\al \be)}(x, \theta^{\pm}) 
=0$. \footnote{An ${\cal N}=3$ superfield $\Theta(x, \theta^{\pm}, 
\theta^0)$ is a function of the bosonic coordinate $x^{\mu}$ and
fermionic coordinates, $\theta^{\pm}$ and $\theta^0$. In the expansion of
$\theta^0$, the component fields are ${\cal N}=2$ superfields $\Phi(x, 
\theta^{\pm})$ where we emphasize the ${\cal N}=2$ superfield by putting 
the dependence of $\theta^{\pm}$ in their arguments. Similarly for 
${\cal N}=3$ superfield, we insert $\theta^0$ dependence explicitly as well
as $\theta^{\pm}$ in 
order 
not to confuse ${\cal N}=2$ one.} 
This $T_{(\al \be)}(x, \theta^{\pm})$ 
is a singlet with respect to 
the flavor group $SU(3)  
$ and its conformal dimension is 2.
Moreover one has one ${\cal N}=2$ 
massless gravitino multiplet characterized by
its conformal dimension $3/2$ and a singlet with respect to the 
flavor group. 
This corresponds to the ${\cal N}=2$ conserved 
current $G^{\al}(x, \theta^{\pm})$ 
relative to the third supersymmetry
charge. It satisfies $D_{\al}^{\pm} G^{\al}(x, \theta^{\pm})=0$. 
The $\theta^0$ independent term of ${\cal N}=3$ 
isospin superfield $\Theta^{\al}(x, 
\theta^{\pm}, \theta^0)$ contains $G^{\al}(x, \theta^{\pm})$  and 
linear term in 
$\theta^0$  has $T_{(\al \be)}(x, \theta^{\pm})$. Therefore one 
can see the conformal dimension of $\theta^0$ is $-1/2$. 

$2)$ One ${\cal N}=3$  massless vector multiplet $\Theta^{i}_j(x, 
\theta^{\pm},\theta^0)$:
$
{\bf 8}, \qquad
 \Delta=1, \qquad J=1
$
 
One can construct $SO(3)_R$ triplet by taking tensor product 
of the isospin doublet in the fundamental representation  of
the flavor group $SU(3)$, $\Theta^i_{J=1/2}(x, \theta^{\pm},\theta^0)$ times 
its conjugate doublet $\Theta_j^{J=1/2}(x, \theta^{\pm}, \theta^0)$.
There exists a conserved vector current, an ${\cal N}=2$ massless vector 
$\Si^i_j(x, \theta^{\pm})$,
to the generator of the flavor
symmetry group  $SU(3)$ through Noether theorem satisfying the conservation
equations $D^{\pm \al} D^{\pm}_{ \al} \Si^i_j(x, \theta^{\pm}) = 0 $. 
This $\Si^i_j(x, \theta^{\pm})$ transforms  in the adjoint  representation
$\bf 8$ of $SU(3)$ flavor group and its conformal dimension is 1.
The remaining two $\theta^0$ independent components 
whose third components of $J$( or $R$-charge) are 1 and $-1$ are
${\cal N}=2$ chiral superfields.

$3)$ One ${\cal N}=3$  massless vector multiplet $\Theta(x, 
\theta^{\pm}, \theta^0)$:
$
{\bf 1}, \qquad
 \Delta=1, \qquad J=1
$

The $\theta^0$ independent terms in this isospin superfield have   
two chiral ${\cal N}=2$ superfields and a linear superfield 
$\Si_{\mbox{betti}}$
satisfying $D^{\pm \al} D^{\pm}_{ \al} 
\Si_{\mbox{betti}}(x, \theta^{\pm}) = 0 $.
It is known that  the Betti current $\Si_{\mbox{betti}}(x, \theta^{\pm})$ 
of 
$N^{0,1,0}$ is obtained and this corresponds to
additional ${\cal N}=2$ massless vector multiplet related to
nonperturbative baryon states carrying Betti charge.

$\bullet$ {\bf ${\cal N}=3$ Short multiplets} 
\cite{termonia,freetal,billoetal}

It is known that 
the dimension of the scalar operator in terms of energy labels,
in the dual SCFT corresponding
$AdS_4 \times N^{0,1,0}$ is
\bea
\Delta = \frac{3}{2} + \frac{1}{2} \sqrt{1 + \frac{m^2}{4}} =
\frac{3}{2} + \frac{1}{2} \sqrt{45 + \frac{E}{4} -6 \sqrt{36 +E}} .
\label{delta}
\eea 
The energy spectrum on $N^{0,1,0}$ exhibits an interesting feature
which is relevant to superconformal algebra and it is given by
\bea
E =  
\frac{16}{3} \left[ 2 \left( M_1^2 +M_2^2 + M_1 M_2 +3M_1 +3M_2 \right) -
3 J(J+1)  \right]
\label{energy}
\eea
where the eigenvalue $E$ is classified by
$SU(3)$ quantum numbers $(M_1, M_2)$ and
$SU(2)$ isospin $J$: $M_1=0,1,2, \cdots$, $M_2=M_1 +3 j$,
$J=j, j+1, \cdots$ and $j \geq 0$. 
The corresponding eigenmodes occur in $( M_1, M_1+3j)$ 
$SU(3)$ representation and   
the angular momentum $J$ $SU(2)$ representation.
According to \cite{termonia,freetal,billoetal}, 
the information on the Laplacian eigenvalues allows us 
to get the spectrum of ${\cal N}=2$ 
hypermultiplets of the theory corresponding to
the chiral operators of the SCFT.
This part of spectrum was given in \cite{termonia,freetal,billoetal} 
and the form of operators
is 
\bea
\mbox{Tr} \Phi_{\mbox{c}} \equiv \mbox{Tr} (U V)^{R},
\qquad 
\left(\bf 1+R \right)^3, \qquad
 \Delta=R, \qquad
J=R \geq 2 
\label{chiral}
\eea
where the flavor $SU(3)$  indices 
are totally symmetrized and the chiral superfield
$\Phi_{\mbox{c}}(x, \theta^{\pm})$ satifies $D_{\al}^{+} 
\Phi_{\mbox{c}}(x, \theta^{\pm}) =0$. 
There exists also other type of chiral superfield $
\mbox{Tr} \Psi_{\mbox{c}}^{\dagger} 
\equiv \mbox{Tr} (\overline{V} \overline{U})^{R}$
with lowest value of third component of $J$.
In the ${\cal N}=2$ theory, it has $R$-charge $-R$. 
The hypermultiplet spectrum in the KK harmonic expansions on $N^{0,1,0}$
agrees with the chiral superfield predicted by the 
conformal gauge theory.
From this, the dimension of $U V$ should be 1 to match the spectrum.
In fact, the conformal weight of a product of
chiral fields equals the sum of the weights of the single components.
This is due to the the relation of $\Delta=R$ satisfied by
chiral superfields and to the additivity of the $R$-charge.

$1)$ One ${\cal N}=3$  short graviton multiplet $\Theta^{\al}
(x, \theta^{\pm},
\theta^0)$:
\bea
\left(\bf 1+R \right)^3, \qquad
 \Delta=R+3/2 , \qquad J=R \geq 1
\label{shortgra}
\eea

Short gravitons of higher isospin can be obtained by
multiplying the $J=0$ massless graviton
 $\Theta^{\al}(x, 
\theta^{\pm}, \theta^0)$ with chiral superfields 
 $\Theta(x, 
\theta^{\pm}, \theta^0)$
of any $J$.
The gauge theory interpretation of this multiplet is obtained by
adding a dimension 2 singlet operator with respect to 
flavor group into the above chiral superfield 
$ \Phi_{\mbox{c}}(x, \theta^{\pm})$ in the ${\cal N}=2$ language.
We consider
$
\mbox{Tr} \Phi_{(\al \be)} 
\equiv \mbox{Tr} \left( T_{(\al \be)}  \Phi_{\mbox{c}} 
\right),$
where $T_{(\al \be)}(x, \theta^{\pm})$ 
is a stress energy tensor 
and $ \Phi_{\mbox{c}}(x, \theta^{\pm})$ is a chiral superfield 
(\ref{chiral}). 
In ${\cal N}=3$ language,  this composite operator
is the coefficient function appearing in the linear $\theta^0$.
Note that the conformal dimension of this composite operator
is given by $\Delta=(R+3/2)+1/2$ which consists of
the conformal dimension of 
$\Theta^{\al}
(x, \theta^{\pm},
\theta^0)$ plus $1/2$.
\footnote{The ${\cal N}=3$ short graviton multiplet
characterized by (\ref{shortgra})
decomposes into ${\cal N}=2$ multiplets \cite{freetal} 
\bea
SD\left(2,\Delta,J \right) \longrightarrow \bigoplus 
\limits_{y=-J}^{J} SD\left(2,\Delta+1/2,y\right)
\oplus\bigoplus\limits_{y=-J}^{J} SD\left(3/2,\Delta,y \right)
\nonu
\eea
where the first element in the left hand side gives us the maximal spin.
So they are 2, 3/2, 1 for graviton, gravitino and vector, respectively.
In the right hand side, $y$ is a $U(1)_R$ charge.
Therefore, ${\cal N}=3$ short graviton multiplet consists of
two short gravitons, $(2R-1)$ long gravitons, two short gravitinos and
$(2R-1)$ long gravitinos in ${\cal N}=2$ language. }
As before we have $D_{\al}^{\pm} T^{ (\al \be)}(x, \theta^{\pm})=0$.
All color indices are symmetrized before taking the contraction.
This composite operator corresponding to ${\cal N}=2$ short graviton 
multiplet
satisfies $D^{+}_{\al} 
\Phi^{(\al \be)}(x, \theta^{\pm}) =0 $.
There exists an operator
$
\mbox{Tr} \Psi_{(\al \be)}^{ \dagger} 
\equiv \mbox{Tr} \left( T_{(\al \be)}  \Psi_{\mbox{c}}^{\dagger} 
\right)$ where 
$ \Psi_{\mbox{c}}^{\dagger}(x, \theta^{\pm})$ is a chiral superfield
and whose $R$-charge in the ${\cal N}=2$ theory is $-R$.
Other component of  gauge theory operator is obtained by
adding a dimension 3/2 singlet operator with respect to 
flavor group into the above chiral superfield 
$ \Phi_{\mbox{c}}(x, \theta^{\pm})$.
That is, 
$
\mbox{Tr} \Phi_{\al } 
\equiv \mbox{Tr} \left( G_{\al }  \Phi_{\mbox{c}} 
\right),$ corresponding to ${\cal N}=2$ short gravitino multiplet
where $G_{\al}(x, \theta^{\pm})$ satisfying  
$D_{\al}^{\pm} G^{\al}(x, \theta^{\pm})=0$
is a conserved current we have seen before. 
There exists an operator
$
\mbox{Tr} \Psi_{\al }^{ \dagger} 
\equiv \mbox{Tr} \left( G_{\al }  \Psi_{\mbox{c}}^{\dagger} 
\right)$ corresponding to other
 ${\cal N}=2$ short gravitino multiplet.
Moreover, according to the decomposition of $OSp(3|4)$
unitary irreducible representation into $OSp(2|4)$ one \cite{freetal},
there exist also
long gravitons and gravitinos that have rational conformal
dimensions.

$2)$ One ${\cal N}=3$  short vector multiplet $\Theta(x, \theta^{\pm},
\theta^0)$:
\bea
\left(\bf 1+R \right)^3, \qquad
 \Delta=R, \qquad J= R \geq 2
\label{shortvec}
\eea

There exist two chiral ${\cal N}=2$ multiplets 
$\mbox{Tr} \Phi_{\mbox{c}}$ and 
$\mbox{Tr} \Psi_{\mbox{c}}^{\dagger}$ 
we have discussed before in this ${\cal N}=3$
short vector multiplet. In addition to them, 
one can construct the following gauge theory object,
based on the decomposition rule of \cite{freetal},
$
\mbox{Tr} \Phi  \equiv \mbox{Tr} \left( \Si_{j}^i \left(UV 
\right)^{R-1} 
\right),$
where $\Si_{j}^i(x, \theta^{\pm})$ 
is a conserved vector current with a singlet 
under the flavor group. The $R$-charge of $\Phi(x, \theta^{\pm})$ 
is $R-1$. 
The Dynkin label of $\Si^i_j(x, \theta^{\pm})$
is $(1,1)$ while those of $(UV)^{R-1}$ is $(R-1,R-1)$.
Moreover we have 
other type of gauge theory object, 
 $
\mbox{Tr} \Psi  \equiv \mbox{Tr} \left( \Si_{j}^i 
\left( \overline{V} \overline{U} \right)^{R-1} 
\right)$ whose $R$-charge is $-(R-1)$.
These composite operators correspond to ${\cal N}=2$ short
vector multiplets.
There exist also ${\cal N}=2$
long vector multiplets whose conformal dimensions are rational.
\footnote{ The ${\cal N}=3$ short vector multiplet by
(\ref{shortvec}) decomposes as follows \cite{freetal}:
\bea
SD\left(1,\Delta,J \right)\longrightarrow\bigoplus 
\limits_{y=-J+1}^{J-1}SD\left(1,\Delta,y\right) 
\oplus SD\left(1/2,\Delta,J\right) 
\oplus SD\left(1/2,\Delta,-J\right).
\nonu
\eea 
So it consists of two chiral multiplets, two short vector multiplets and
$(2R-1)$ long vectors.}

$3)$ One ${\cal N}=3$ short gravitino multiplet $\Theta^{\al}(x, 
\theta^{\pm}, \theta^0)$:
\bea 
\frac{1}{2} \left(\bf 1+R \right)\left( \bf 4+R \right)\left( 
\bf 5+2R \right), \qquad
 \Delta=R+2, \qquad J=R+1, \qquad R \geq 0
\label{shortgravi}
\eea

Short gravitinos of higher isospin can be obtained by
multiplying the $J=1$
short gravitino $\Theta^{\al}(x, 
\theta^{\pm}, \theta^0)$
with chiral superfields   $\Theta(x, 
\theta^{\pm}, \theta^0)$ of 
any $J$.
From the above $SU(3)$ representation of this multiplet,
one can consider
$
\mbox{Tr} \Phi^{+ (ijk)}_{\al}  
\equiv \mbox{Tr} \left( G^{+ (ijk)}_{\al}  \Phi_{\mbox{c}} 
\right),$
where $G^{+ (ijk)}_{\al}(x, \theta^{\pm})$ 
is a conserved vector current with conformal dimension $5/2$ 
transforming in the $ {\Box\!\Box\!\Box}^\ast$ representation
of $SU(3)$ flavor group. 
\footnote{In terms of ${\cal N}=2$ hypermultiplets,   
we have $G^{+ (ijk)}_{\al}=
 f^{lm(i} U^j U^{k)} \left( \overline {U_l} D^{+}_{\al} \overline{U_m} -  
\overline{U_m}
D^{+}_{\al} \overline{U_l} \right)$ \cite{billoetal}. }
Notice that the conformal dimension 
of this composite operator is made of the one of
$\Theta^{\al}(x, 
\theta^{\pm}, \theta^0)$ plus $1/2$. 
\footnote{
${\cal N}=3$ short gravitino multiplet with (\ref{shortgravi})
breaks into the following scheme \cite{freetal}
\bea
SD\left(3/2,\Delta,J\right)\longrightarrow\bigoplus 
\limits_{y=-J}^{J}SD\left(3/2,\Delta+1/2,y\right) 
\oplus\bigoplus\limits_{y=-J}^{J}SD\left(1,\Delta,y\right). 
\nonu
\eea
Therefore, it consists of two short gravitinos, 
two short vectors, $(2R+1)$ long gravitinos and
$(2R+1)$ long vectors in terms of ${\cal N}=2$ superfields.}
This is one of the two ${\cal N}=2$ short gravitino 
multiplets. The other is 
the following gauge theory object $
\mbox{Tr} \Psi^{- (ijk),\dagger}_{\al}  
\equiv \mbox{Tr} \left( G^{- (ijk)}_{\al}  \Psi_{\mbox{c}}^{\dagger} 
\right)$ where $R$-charge is $-(R+1)$. 
There exist two ${\cal N}=2$ short vector multiplets:
one is realized by 
$
\mbox{Tr} \Phi^{+ (ijk)}  
\equiv \mbox{Tr} \left( \Si^{+ (ijk)}  \Phi_{\mbox{c}} 
\right),$
where $\Si^{+ (ijk)}(x, \theta^{\pm})$ 
is a conserved vector current of dimension 2 
transforming as the $ {\Box\!\Box\!\Box}^\ast$ representation of 
flavor group and the other by
$
\mbox{Tr} \Psi^{- (ijk),\dagger}  
\equiv \mbox{Tr} \left( \Si^{- (ijk)}  \Psi_{\mbox{c}}^{\dagger} 
\right)$. 
\footnote{ In terms of two chiral fields we have explicit forms:
 $G^{- (ijk)}_{\al}= -
 f^{lm(i} {\overline{V}}^j {\overline{V}}^{k)} \left( V_l D^{-}_{\al} V_m -  
V_m 
D^{-}_{\al} V_l \right)$ and $
\Si^{-(ijk)} = - f^{lm(i} \overline{V}^j \overline{V}^{k)} \left(
 V_l \overline{U}_m - \overline{U}_m V_l \right)$.
For $\Si^{+(ijk)}$, see the equation (\ref{Si}). } 
Rational 
long gravitino and vector multiplets are present in this case.

$\bullet$ {\bf ${\cal N}=3$ Long multiplets} 
\cite{termonia,freetal,billoetal}

Although the dimensions of nonchiral operators are in general irrational,
there exist special integer values of $j$ such that
for $M_1=R, M_2=R+3j$ and $J=R+j$,
one can see the 
condition, 
$
j \left(j-1 \right)=0
$
make $\sqrt{36+E}$ be equal to
$2 \left(2R+4j +3 \right)$(See also \cite{ahnplb,ahn02,ahn02-1,ahn02-2}).
We consider here maximal $J=R+j$ because from the energy eigenvalues
$E$ (\ref{energy}) this will give us the lowest conformal dimension.
Also one can generalize this analysis for nonmaximal case, $J=R+j-k$
where $k$ is some positive integers. 
Furthermore in order to make 
the dimension be rational(their conformal dimensions are protected), 
$45 + E/4 -6 \sqrt{36 +E}$  in (\ref{delta}) should be square of 
something. 
Therefore we  have $\De=R+ 2j $. 
This is true if we are describing states with finite $\Delta$ and $R$.
Since we are studying the scaling limit $\Delta, R \rightarrow \infty$,
we have to modify the above analysis.
The 
energy eigenvalue of the Laplacian on $N^{0,1,0}$ for the supergravity mode
(\ref{energy})
takes the form 
\bea
E = 4 \left( R^2 + 4 j R + 5 j^2 + 3 R + 5 j \right) .    
\label{energyr}
\eea
One can show that the conformal weight of the ${\cal N}=3$ long multiplets 
below becomes rational if $j=0$ or $j=1$.  

$1)$ One ${\cal N}=3$ long gravitino multiplet:
\bea
{\bf \frac{(1+R)(1+R+3j)(2+ 2R+3j)}{2}}, 
\;
\Delta= \pm \frac{3}{2} +\frac{1}{4} \sqrt{E+36}, \;
j \leq J \leq j+R,
\; R \geq 0, \; j \geq 2
\label{longgra}
\eea

For finite $R$ with rational dimension, 
after inserting the $E$ into the above, 
we will arrive at the relation with same constraint:
\bea
\Delta-R = 2j   +{\cal O} \left(\frac{1}{R} \right), \qquad
\Delta-R = 3+ 2j   +{\cal O} \left(\frac{1}{R} \right).
\label{delta2}
\eea
So the constraint $j=0$ or $j=1$ is not relevant in the
subsector of the Hilbert space we are interested in. 
Candidates for such states in the gauge theory side are given in terms of
semi-conserved superfields \cite{ahn02,ahn02-1,ahn02-2}.
According to the observation  \cite{freetal} of
the decomposition of ${\cal N}=3$ into ${\cal N}=2$ of the multiplets,
the ${\cal N}=3$ long gravitino decomposes into 
various ${\cal N}=2$ long gravitino mulptiplets whose conformal dimensions
are greater than the above (\ref{delta2})($\Delta -R = 2j+1/2$ or 
$2j+7/2$) and ${\cal N}=2$
long vector multiplets whose conformal dimensions are $\Delta-R=2j, 
2j+1, 2j+3$ or $2j+4$.
\footnote{${\cal N}=3$ long gravitino multiplet characterized 
by (\ref{longgra}) decomposes into its ${\cal N}=2$ components fields 
\cite{freetal}
\bea
SD\left(3/2,\Delta,J\right)\longrightarrow\bigoplus 
\limits_{y=-J}^{J}SD\left(3/2,\Delta+1/2,y\right) 
\oplus\bigoplus\limits_{y=-J}^{J}SD\left(1,\Delta+1,y\right)
\oplus\bigoplus\limits_{y=-J}^{J}SD\left(1,E_0,y\right), \;
\Delta >J+1.
\nonu
\eea
So it consists of $(2R+1)$ long gravitinos and $2(2R+1)$ long vectors. } 
Although they are not chiral primaries, their
conformal dimensions are protected. The ones we are interested in take the
following form,
\bea
\mbox{Tr} \Phi_{\mbox{s.c.}} \equiv
\mbox{Tr} \left[  \left( \Si^{+(klm)} \right)^{j}  
  \left(U V \right)^{R} \right], \qquad \Delta-R =2j, \qquad
J=j+R
\label{semi}
\eea   
where the scalar superfields $\Si^{+(klm)}(x,\theta^{\pm})$ transform 
in the $ {\Box\!\Box\!\Box}^\ast$ 
representation of flavor group $SU(3)$ and satisfy
$D^{\pm \al} D^{\pm}_{\al} \Si^{+(klm)}(x,\theta^{\pm}) =
0 $ with conformal dimension 2. 
In this case, also we have $D^{+ \al} D^{+}_{\al} \Phi{
\mbox{s.c.}}(x, \theta^{\pm})=0$. 
Since the singleton superfields $U^{i,a}_{b}$
carry  index $a$ in the ${\bf N}$ of $SU(N)_1$ and 
index $b$ in the ${\bf \overline{N}}$ of the $SU(N)_2$,
the fields $V_{j,a}^{b}$
carry index $a$ in the $ {\bf \overline{N}}$ of $SU(N)_1$ and 
index $b$ in the  ${\bf N}$ of the $SU(N)_2$,
one can construct the following conserved flavor
current  \cite{billoetal}
\bea
 \Si^{+(klm)} & = & f^{ij(k} U^l U^{m)} 
\left( V_i \overline{U_j} -V_j \overline{U_i} \right)
\label{Si}
\eea
where the color indices are contracted in the right hand side.
Note that the conformal dimension of these currents is not
the one of naive sum of $U$ and $\overline{U}$ and 
$V$ and $\overline{V}$. 
As we discussed in the last section, supergravity theory in $AdS_4 \times
N^{0,1,0}$ acquires an enhanced ${\cal N}=8$ superconformal symmetry in the
Penrose limit. This implies that the spectrum of the gauge theory 
operators in this subsector should fall into
${\cal N}=8$ multiplets. We expect that both the chiral primary
fields of the form  $\mbox{Tr} (U V)^{R}$ (\ref{chiral}) and the 
semi-conserved multiplets
of the form (\ref{semi}) combine into  ${\cal N}=8$ multiplets in the 
limit. Note that for finite $R$, the semi-conserved multiplets
should obey the condition $j=0$ or $j=1$  in order for them to
possess rational conformal weights. 

$\bullet$ ${\cal N}=2$ long vectors in ${\cal N}=3$ long gravitino:
$\Delta-R= 2j, 2j+1, 2j+3, 2j+4; J=j+R$

According to the above analysis, the semi-conserved superfield (\ref{semi})
in the gauge theory side gives rise to 
${\cal N}=2$ long vector multiplet whose conformal
dimension $\Delta-R=2j$. 
To produce a higher dimension operator with same $SU(3)$ representation and
$SU(2)$ isospin representation, one has to multiply a dimension 1 operator 
under
a singlet with respect to flavor group with
a semi-conserved superfield. The candidate for this is a Betti current whose
third component of $J$ is zero and the dimension is 1. So we obtain the 
following
gauge theory object
$\mbox{Tr} \left( \Si_{\mbox{betti}}  \Phi_{\mbox{s.c.}}\right)$
corresponding to ${\cal N}=2$ long vector multiplets whose the conformal 
dimension
is $\Delta-R=2j+1$.
In order to construct a gauge theory operator corresponding to
an ${\cal N}=2$ long vector multiplet whose conformal dimension $\Delta-R=
2j +3$, we make 
$\mbox{Tr} \left( T_{(\al \be)}^{0}  \Phi_{\mbox{s.c.}} 
\right)$ where dimension 3 opeerator 
$ T_{(\al \be)}^{0}(x, \theta^{\pm})$ with $R$-charge 0 
appears in the linear $\theta^0$ term of $\Theta^{\al}(x, 
\theta^{\pm}, \theta^0)$ with $J=1$. Finally for
last gauge theory operator in this series, by taking the quadratic
expression of dimension 2 stress-energy tensor with a semi-conserved
field one can make the following combination 
$\mbox{Tr} \left( T_{(\al \be)}^2  
\Phi_{\mbox{s.c.}}\right)$ corresponding to ${\cal N}=2$ long
vector multiplet whose conformal dimension is $\Delta-R=2j+4$.  

$\bullet$ ${\cal N}=2$ long gravitinos in ${\cal N}=3$ long gravitino:
$\Delta-R= 2j+1/2, 2j+7/2; J=j+R$

Since we do not have any dimension 1/2 operator whose R-charge is 
zero and a singlet 
under a flavor group, we have to replace a single $\Si^{+(klm)}$  
among the product  of $(\Si^{+(klm)})^j$
with dimension 3 operator $G^{+(npq)}_{\al}$ by noting that the conformal 
dimension
can be decomposed as $\Delta-R=2(j-1) + 5/2$. Therefore one can 
construct
the following object which satisfies the right representations: $  
\mbox{Tr} \left[ (\Si^{+(klm)})^{j-1} G^{+(npq)}_{\al}(UV)^R \right]
$.
Next, for the second long gravitino, one can multiply dimension 5/2
operator which
is neutral under the isospin to
dimension $(R+2j+1/2)$ operator:
$\mbox{Tr} \left(  (\Si^{+(klm)})^{j-1} G^{+(npq)}_{\al} (UV)^R 
T_{(\be \ga)}^{0} 
\right)$.

$2)$ One ${\cal N}=3$  long graviton multiplet :
\bea
{\bf \frac{(1+R)(1+R+3j)(2+ 2R+3j)}{2}}, 
\;
\Delta= \frac{1}{4} \sqrt{E+36}, \; j \leq J \leq j+R,
\; R \geq 0, \; j \geq 2
\label{longgraviton}
\eea

The combination of 
$\Delta-R$ is given by   
\bea
\Delta-R = \frac{3}{2} +2j  + {\cal O}(\frac{1}{R})
\nonu
\eea
where the right hand side is definitely rational and they are integers.
According to the observation \cite{freetal} of
the decomposition of ${\cal N}=3$ into ${\cal N}=2$ of the multiplets,
the ${\cal N}=3$ long graviton multiplet decomposes into 
various ${\cal N}=2$ long graviton mulptiplets whose conformal dimension
is $\Delta -R = 2j+2$,
long gravitino multiplets where the conformal dimensions are 
$\Delta-R=2j+3/2$ or $2j+5/2$
and
long vector multiplets whose conformal dimension is $\Delta-R=2j+2$.
\footnote{ ${\cal N}=3$  long graviton multiplet with the condition
(\ref{longgraviton}) breaks into \cite{freetal}
\bea
SD\left(2,\Delta,J\right)
& \longrightarrow & \bigoplus\limits_{y=-J}^{J}SD 
\left(2,\Delta+1/2,y\right) 
\oplus\bigoplus\limits_{y=-J}^{J}SD\left(3/2,\Delta,y\vert 2\right) 
\oplus\bigoplus\limits_{y=-J}^{J}SD\left(3/2,\Delta+1,y\right) \nonu \\
&& \oplus 
\bigoplus\limits_{y=-J}^{J}SD\left(1,\Delta+1/2,y\right), \qquad
\Delta > J +3/2. 
\nonu
\eea
It consists of $(2R+1)$ long gravitons, $2(2R+1)$ long gravitinos, 
and $(2R+1)$ long vectors.}

$\bullet$ ${\cal N}=2$ long gravitons in ${\cal N}=3$ long graviton:
$\Delta-R=2j+2;J=j+R$

By adding conformal dimension 2 singlet operator of neutral under 
the isospin 
to a semi-conserved current, we can construct
a gauge theory operator corresponding to ${\cal N}=2$ 
long graviton multiplet,
$\mbox{Tr} \left( T_{(\al \be)}  \Phi_{\mbox{s.c.}}
\right)$.

$\bullet$ ${\cal N}=2$ long gravitinos in ${\cal N}=3$ long graviton:
$\Delta-R = 2j+3/2, 2j+5/2;J=j+R$

The gauge theory interpretation of this multiplet is
quite simple. If we take  a semi-conserved current 
$\Phi_{\mbox{s.c.}}(x, \theta^{\pm})$ defined in (\ref{semi})
and multiply it by a conserved current  
$G_{\al}(x, \theta^{\pm})$ that is a singlet with respect to the
flavor 
group with conformal dimension $3/2$ and is neutral under the isospin,
namely 
$
\mbox{Tr} \left( G_{\al}  \Phi_{\mbox{s.c.}} 
\right)$,
we reproduce the right $OSp(2|4) \times SU(3) $
representations of the ${\cal N}=2$ long gravitino multiplet:
$\Delta-R=2j+3/2$.
Similarly, 
one can construct the following 
gauge theory operator 
$
\mbox{Tr} \left( \Si_{\mbox{betti}} G_{\al}  \Phi_{\mbox{s.c.}} 
\right)$
whose conformal dimension is $\Delta-R=2j+5/2$.
 
$\bullet$ ${\cal N}=2$ long vectors in ${\cal N}=3$ long graviton:
$\Delta-R=2j+2;J=j+R$

One can think of the following combination for this multiplet
 $\mbox{Tr} \left( \Si_{\mbox{betti}}^2  \Phi_{\mbox{s.c.}}\right)$.

$3)$ One ${\cal N}=3$  long graviton multiplet :
\bea
{\bf \frac{(1+R)(4+R)(5+ 2R)}{2}}, 
\qquad
\Delta= \frac{1}{4} \sqrt{E+36}, \qquad
1 \leq J \leq 1+R,
\;\; R \geq 0
\nonu
\eea

The combination of $\Delta -R$ with Penrose limit 
$R \rightarrow \infty$ in the gauge theory side becomes
$
\Delta-R = 7/2   +{\cal O} \left(\frac{1}{R} \right).
$
By recognizing that this is a particular case of $j=1$ in previous one, 
we can  construct corresponding gauge theory operator 
$
 \mbox{Tr} \left( G_{\al} \Si^{+(klm)} (UV)^R \right)$
corresponding to
${\cal N}=2$ long gravitino multiplet. For ${\cal N}=2$ long graviton
multiplet, we have
the following gauge theory operator  $
 \mbox{Tr} \left( \Si_{\mbox{betti}}^2   \Si^{+(klm)} (UV)^R \right)$ or
 $
 \mbox{Tr} \left( T_{(\al \be)}   \Si^{+(klm)} (UV)^R \right)$ and
for long gravitino of dimension 9/2, there exists
an operator 
$\mbox{Tr} \left(  \Si_{\mbox{betti}} G_{\al} \Si^{+(klm)} (UV)^R  
\right)$.

$4)$ One ${\cal N}=3$  long gravitino multiplet:
\bea
{\bf \frac{(1+R)(4+R)(5+ 2R)}{2}}, 
\qquad
\Delta= \pm \frac{3}{2} +\frac{1}{4} \sqrt{E+36},
\qquad
1 \leq J \leq 
1+R [R], \qquad 
R \geq 0
\nonu
\eea
where it is understood that the maximum value of $J$ of
gravitino with larger conformal dimension is $1+R$ while
the one with smaller conformal dimension is $R$.
In this case, for smaller conformal dimension
we have to take $J=R+j-1$ not $R+j$ with $j=1$.
With the same value of $M_1$ and $M_2$, we get
the energy eigenvalue  of the Lapacian on $N^{0,1,0}$
and arrive at one higher conformal dimension. However for
larger conformal dimension, 
we get $\Delta -R$  from  the one in (\ref{delta2})
by setting $j=1$. Combining these we get 
$
\Delta-R = 3  +{\cal O} \left(\frac{1}{R} \right)$ and $
\Delta-R = 5  +{\cal O} \left(\frac{1}{R} \right)
$.
One can describe a gauge theory operator by taking
a single conserved current $ \Si^{+(klm)}(x, \theta^{\pm})$
from a semi-conserved 
current $\Phi_{\mbox{s.c.}}(x, \theta)$
in order to match with 
 the conformal dimension. That is, one obtains  
the following new gauge theory operators
related to ${\cal N}=2$ long vectors with $\Delta-R=3, 4$
are given by 
$
\mbox{Tr} \left( \Si_{\mbox{betti}} \Si^{0(klm)} (UV)^R  \right)
$ and $ 
 \mbox{Tr} \left( {\Si_{\mbox{betti}}}^2   
\Si^{0(klm)} (UV)^R \right) 
$ repectively
by imposing the condition that 
the highest $J$ value is $J=R$ into the above operators. 
\footnote{The neutral $\Si^{0(ijk)}$ is given in terms of
chiral fields: $\Si^{0(ijk)}=\sqrt{2} i f^{lm(i} U^j \overline{V}^{k)}
\left( V_l \overline{U_m} -V_m \overline{U_l} \right)$. }
When $\Delta-R=5,6$
one can read off corresponding gauge theory objects from the expressions
in ${\cal N}=2$ long vectors with general $j$  by putting $j=1$
we have discussed before.
Similarly 
gauge theory operator
related to ${\cal N}=2$ long gravitinos with $\Delta-R=7/2$
is given by
$
\mbox{Tr} \left( \Si_{\mbox{betti}} G^{0(npq)}_{\al}(UV)^R \right) 
$
where we also put the right isospin charge count $J=R$.

$5)$ One ${\cal N}=3$  long graviton multiplet:
\bea
{\bf (1+R)^3}, 
\qquad
\Delta= \frac{1}{4} \sqrt{E+36},
\qquad
0 \leq J \leq R-1,
\qquad R \geq 0
\nonu
\eea

The combination of $\Delta -R$ with Penrose limit 
$R \rightarrow \infty$ in the gauge theory side becomes
$
\Delta-R = 5/2  +{\cal O} \left(\frac{1}{R} \right)
$ by taking same $M_1$ and $M_2$ values with $J=R-1$.  
The gauge theory operators
related to ${\cal N}=2$ long gravitinos with $\Delta-R=5/2, 3, 7/2$
are given by 
$
 \mbox{Tr} \left( \Si_{\mbox{betti}} \Si^i_j G_{\al}  (UV)^{R-1} \right)$, 
$ 
\mbox{Tr} \left( \Si_{\mbox{betti}} \Si^i_j T_{(\al \be)}  (UV)^{R-1} 
\right)$ and 
$
 \mbox{Tr} \left( {\Si_{\mbox{betti}}}^2 
\Si^i_j G_{\al}  (UV)^{R-1} \right)$.

$6)$ One ${\cal N}=3$  long gravitino multiplet:
\bea
{\bf (1+R)^3}, 
\qquad
\Delta= \pm \frac{3}{2} +\frac{1}{4} \sqrt{E+36},
\qquad
0 \leq J \leq R[R-1],
\qquad R \geq 0
\nonu
\eea
In this case, the maximum value of $J$ is $R$ for larger conformal
dimension and the one for smaller conformal dimension is $R-1$. 
The combination of $\Delta -R$ with Penrose limit 
$R \rightarrow \infty$ in the gauge theory side becomes
$
\Delta-R =  1+ {\cal O} \left(\frac{1}{R} \right)$ and
$ 
\Delta-R =  3+ {\cal O} \left(\frac{1}{R} \right).
$ 
 Note that from the isospin representation of $\Delta-R=1$ in the 
${\cal N}=3$ language,
the highest $J$ value is $J=R-1$. Therefore we have to take into 
account this fact. This is the only
difference between general $j$ case and $j=0$ case here. So we 
imposed this condition 
for first two cases below.  
The new gauge theory operators
related to ${\cal N}=2$ long vectors with $\Delta-R=1,2$
with right isospin counting
are $
\mbox{Tr} \left( \Si_{\mbox{betti}} \Si_{j}^i     (UV)^{R-1}  \right)$
and $
\mbox{Tr} \left( {\Si_{\mbox{betti}}}^2 \Si_{j}^i    
(UV)^{R-1} \right)$.
For $ \Delta-R=3,4$ we get similar expressions for previous case with $j=0$.
Finally,
the gauge theory operator
related to ${\cal N}=2$ long gravitino with $\Delta-R=3/2$
is
$
\mbox{Tr} \left(  \Si^i_j G_{\al} (UV)^{R-1} \right)
$.

\section{Conclusion}
We described an explicit example of an ${\cal N}=3$
superconformal field theory that has a subsector of the Hilbert space
with enhanced ${\cal N}=8$ superconformal symmetry, in the large $N$ limit
from the study of $AdS_4 \times N^{0,1,0}$.
The pp-wave geometry in the scaling limit produced to 
the maximally ${\cal N}=8$  supersymmetric pp-wave
solution of $AdS_4 \times \S^7$.
The result of this paper shares common characteristic feature 
of previous cases of $AdS_4 \times Q^{1,1,1}$ \cite{ahn02},
$AdS_4 \times M^{1,1,1}$ \cite{ahn02-1} and $AdS_4 \times V_{5,2}$ 
\cite{ahn02-2}.
This subsector of gauge theory is achieved by Penrose limit
which  constrains strictly the states of the gauge theory to those
whose conformal dimension and $R$ charge diverge in the large $N$ limit
but possesses finite value $\Delta-R$.
We predicted for the spectrum of $\Delta-R$ of the ${\cal N}=3$
superconformal field theory and proposed how the exicited states in the
supergravity correspond to
gauge theory operators. In particular, both the chiral multiplets 
(\ref{chiral}) and
semi-conserved multiplets (\ref{semi}) should
combine into ${\cal N}=8$ chiral multiplets.
It would be interesting to find out a Penrose limit of other types 
\cite{ahnetal} of M-theory
compactification,  along the line of \cite{warneretal}. 
These examples have different supersymmetries and the 
structures of four-form field strengths are more complicated than what we 
have discussed.         
      
\vskip1cm
$\bf Acknowledgements$

This research was supported 
by 
grant 2000-1-11200-001-3 from the Basic Research Program of the Korea
Science $\&$ Engineering Foundation.

\end{document}